\documentclass[aps,prl,preprint,superscriptaddress]{revtex4}
\usepackage{graphicx}
\usepackage{dcolumn}
\usepackage{bm}
\usepackage[usenames]{color}
\usepackage{amsmath}
\usepackage{amssymb}
\begin{document}
\title{
Parity Violation in a Single Domain of Spin-Triplet Sr$_2$RuO$_4$ Superconductors\\
}
\author{H. Nobukane}
\affiliation{Department of Applied Physics, Hokkaido University,
Sapporo 060-8628, Japan}
\author{K. Inagaki}
\affiliation{Department of Applied Physics, Hokkaido University,
Sapporo 060-8628, Japan}
\affiliation{The 21st Century COE Program ``Topological Science and Technology'' from the MEXT of Japan, Hokkaido University, Sapporo 060-8628, Japan}
\author{K. Ichimura}
\affiliation{Department of Applied Physics, Hokkaido University,
Sapporo 060-8628, Japan}
\affiliation{The 21st Century COE Program ``Topological Science and Technology'' from the MEXT of Japan, Hokkaido University, Sapporo 060-8628, Japan}
\author{K. Yamaya}
\affiliation{Department of Applied Physics, Hokkaido University,
Sapporo 060-8628, Japan}
\affiliation{The 21st Century COE Program ``Topological Science and Technology'' from the MEXT of Japan, Hokkaido University, Sapporo 060-8628, Japan}
\author{S. Takayanagi}
\affiliation{Department of Applied Physics, Hokkaido University,
Sapporo 060-8628, Japan}
\affiliation{The 21st Century COE Program ``Topological Science and Technology'' from the MEXT of Japan, Hokkaido University, Sapporo 060-8628, Japan}
\author{I. Kawasaki}
\affiliation{Department of Physics, Hokkaido University,
Sapporo 060-0810, Japan}
\author{K. Tenya}
\affiliation{Department of Education, Shinshu University,
Nagano 380-8544, Japan}
\author{H. Amitsuka}
\affiliation{The 21st Century COE Program ``Topological Science and Technology'' from the MEXT of Japan, Hokkaido University, Sapporo 060-8628, Japan}
\affiliation{Department of Physics, Hokkaido University,
Sapporo 060-0810, Japan}
\author{K. Konno}
\affiliation{Department of Applied Physics, Hokkaido University,
Sapporo 060-8628, Japan}
\affiliation{The 21st Century COE Program ``Topological Science and Technology'' from the MEXT of Japan, Hokkaido University, Sapporo 060-8628, Japan}
\author{Y. Asano}
\affiliation{Department of Applied Physics, Hokkaido University,
Sapporo 060-8628, Japan}
\affiliation{The 21st Century COE Program ``Topological Science and Technology'' from the MEXT of Japan, Hokkaido University, Sapporo 060-8628, Japan}
\author{S. Tanda}
\affiliation{Department of Applied Physics, Hokkaido University,
Sapporo 060-8628, Japan}
\affiliation{The 21st Century COE Program ``Topological Science and Technology'' from the MEXT of Japan, Hokkaido University, Sapporo 060-8628, Japan}
\date{\today}
\begin{abstract}
We observed an unconventional parity-violating vortex in single domain Sr$_2$RuO$_4$ single crystals using a transport measurement.
The current-voltage characteristics of submicron Sr$_2$RuO$_4$ shows that the induced voltage has anomalous components which are \textit{even} function of the bias current.
The results may suggest that the vortex itself has a helical internal structure characterized by a Hopf invariant (a topological invariant).
We also discuss that the hydrodynamics of such a helical vortex causes the parity violation to retain the topological invariant.

\end{abstract}

\maketitle
A quantized vortex is a topological matter in superconductors and superfluids.
In metallic superconductors, the Abrikosov vortex is characterized only by an integer 
winding number of phase. 
In unconventional superconductors, the superfluid of $^3$He and the spinor cold atoms, 
internal degrees of freedom of the order parameter enrich the variety of vortices~\cite{sigrist,volovik,ho}. 
Although a number of theoretical studies have predicted the existence of 
such unconventional vortices, experimental confirmation of them is still limited to
several studies such as NMR spectroscopy of $^3$He and imaging of spinor cold 
atoms~\cite{Ru,spindomain}. 
Here a Cooper pair in spin-triplet superconductors has electric charge $2e$.
Thus one can resolve the dynamics of unconventional vortices through electric
transport. 
We will address this issue in the present Letter. 

Sr$_2$RuO$_4$~\cite{RMP} is a promising candidate of spin-triplet 
chiral-$p$ superconductor, (i.e., spin $S=1$ and orbital angular momentum $L=1$). 
Since two states belonging to different chirality degenerate 
in the ground state, bulk Sr$_2$RuO$_4$
is considered to have chiral domain structures. 
Transport properties have been studied in relation to Josephson interferometry 
using bulk Sr$_2$RuO$_4$ crystals to determine the symmetry of Cooper pairs 
and measure the dynamics of chiral domains~\cite{odd,domain}. 
These experimental data on bulk  Sr$_2$RuO$_4$ should be considered as a 
result of ensemble averaging over possible chiral domain configurations. 
Thus we need a small enough sample of Sr$_2$RuO$_4$ 
rather than the domain size to study peculiar phenomena to a single chiral domain such as dynamics of a single chiral domain, spin supercurrent, and unconventional vortices~\cite{asano,babaev,chung}.
Transport measurements, however, have never been carried out yet in a single domain 
because it is also difficult to attach electrical contacts to submicron 
Sr$_2$RuO$_4$ crystals. 

In this Letter, we will report an anomalous property of current-voltage ($I-V$) characteristics in the single domain of Sr$_2$RuO$_4$.
The creation of vortices gives the finite resistivity even when a temperature
is well below the superconducting transition temperature. 
In four-terminal measurements, the induced voltage $V$ is usually an \textit{odd} function of the bias current $I$. Namely, $V$ changes its sign when we flip the direction of current to the opposite direction, which implies the parity conservation~\cite{parity}. 
However, we find in submicron Sr$_2$RuO$_4$ samples that 
$V$ has anomalous components which are \textit{even} function of $I$.
The existence of the anomalous components means that positive voltage is detected regardless 
of the current direction and suggests the the violation of parity~\cite{parity}.
To understand the nature of the anomalous $I-V$ characteristics, we consider a simple model of vortex which has a helical internal structure characterized by a Hopf invariant.
We also show that hydrodynamics of such a helical vortex violates the parity to retain the topological invariant.   

To obtain submicron Sr$_2$RuO$_4$ single crystals, we synthesized Sr$_2$RuO$_4$ crystals with a solid phase reaction and then determined the crystal structure of Sr$_2$RuO$_4$ and the concentration of impurities. 
We prepared SrCO$_3$ and RuO$_2$ (both 99.9 \%, Kojundo Chem.) powders. 
The mixed powder was then heated at 990 $^\circ$C for 60 hours. 
The mixture cooled gradually from 990 $^\circ$C to 450 $^\circ$C over six hours.
The samples were kept at 450 $^\circ$C for 12 hours to introduce oxide into the crystals and then cooled down slowly at room temperature. 
The structure of the Sr$_2$RuO$_4$ crystals was analyzed by using x-ray power diffraction (Rigaku Diffractometer RINT 2200HK) with Cu $K$$\alpha$ radiation. The observed peaks fitted a body-centered tetragonal unit cell of the K$_2$NiF$_4$ type with lattice constants $a=b=$ 3.867 ($\pm$0.004) \AA\ and $c=$ 12.73 ($\pm$0.01) \AA\ \cite{x-ray}.
The result of secondary ion-microprobe mass spectrometry (SIMS) shows the concentration of the Al in the sample is less than 100 ppm, while the superconductivity of Sr$_2$RuO$_4$ is destroyed by nonmagnetic impurities \cite{impurity}.

We selected submicron Sr$_2$RuO$_4$ single crystals from the results of chemical composition and crystallinity \cite{nobukane}.  
The samples were dispersed in dichloroethane by sonication and deposited on an oxidized Si substrate. 
We found typical samples of about 50 nm $\sim$ 500 $\mu$m. 
Energy dispersion spectroscopy (EDS;EX-64175 JMU, JEOL) was used to determine the components of the submicron samples on the substrate. 
The molar fraction of the Sr and Ru elements was 2 : 1. 
We also confirmed that the dispersed crystals had no boundaries nor ruthenium inclusions on the sample surface by observing the crystal orientation using the electron backscatter diffraction pattern (EBSP; OIM TSL \cite{EBSP}). 

\begin{figure}[t]
\begin{center}
\includegraphics[width=0.6\linewidth]{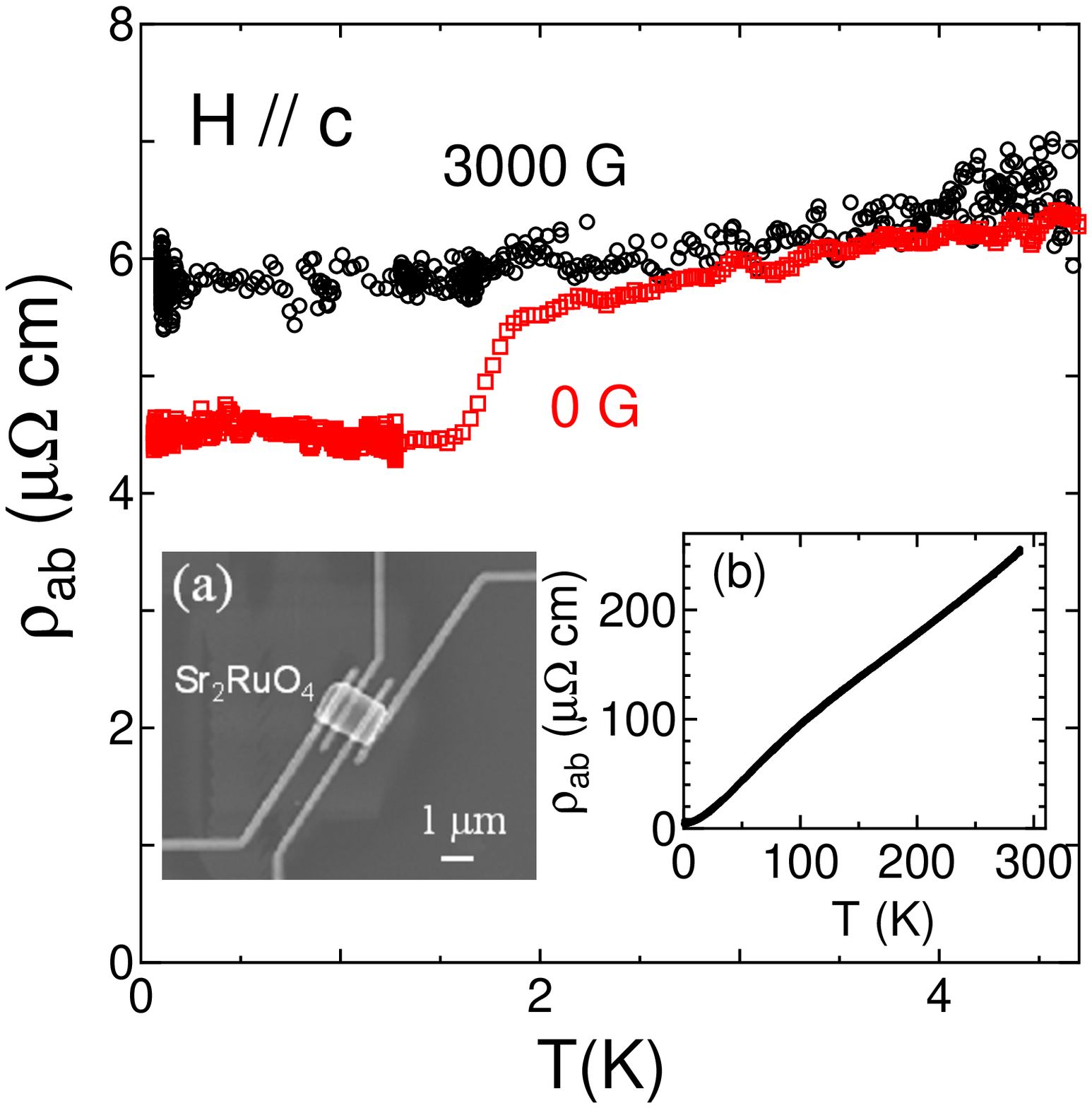}
\caption{
Temperature dependence of resistivity of submicron Sr$_2$RuO$_4$ in zero magnetic field (0 G) 
and in a magnetic field ($H =$ 3000 G) applied parallel to $c$ axis. 
Flat tail resistivity can be seen at low temperatures below $T_c$ = 1.69 K. 
The inset (a) shows a micrograph of a submicron Sr$_2$RuO$_4$ single crystal connected to gold electrodes.
The inset (b) displays temperature dependence of the resistivity in 
the $ab$ plane from room temperature down to 4.2 K }
\label{resistivity}
\end{center}
\end{figure}

On the analysed Sr$_2$RuO$_4$, we fabricated gold electrodes using overlay electron beam lithography. 
The inset (a) in Fig. \ref{resistivity} shows a micrograph of our samples. 
The sample size is 2.50 $\mu$m $\times$ 1.88 $\mu$m $\times$ 0.10 $\mu$m. 
The sample electrode spacing is 0.63 $\mu$m. 
Since the fabricated sample surface may have the insulator surface of the layer crystals and the residual resist between the sample and the gold electrodes, it is difficult to form electrical contact.
Therefore we performed a welding using an electron beam irradiation \cite{inagaki}. 
We heated each electrode on the sample for 15 s with a beam current irradiation of 2 $\times$ 10$^{-7}$ A. 
As the result, we succeeded in greatly reducing the contact resistance below 10 $\Omega$ at room temperature. 

The measurements were carried out in a dilution refrigerator (Kelvinox, Oxford) with a base temperature of 60 mK.
All measurement leads were shielded. 
The lead lines were equipped with low pass RC filters (R = 1 k$\Omega$, C = 22 nF). 
In the DC measurements, a bias current was supplied by a precise current source (6220, Keithley) and the voltage was measured with a nanovoltmeter (182, Keithley) by four-terminal measurements. 
  
We measured the temperature dependence of the resistivity in the submicron Sr$_2$RuO$_4$. 
The inset (b) of Fig. \ref{resistivity} represents temperature dependence of the resistivity in $ab$ plane from room temperature down to 4.2 K.
Figure \ref{resistivity} shows the resistivity $\rho_{ab}$(4K) = 6.0 $\mu\Omega $ cm.
This value is larger than the bulk resistivity by about three times \cite{impurity}. 
We estimated the resistivity $\rho_{ab}$ from the sample size.
Since Sr$_2$RuO$_4$ have anisotropic resistivity $\rho_{ab}$ $\approx$ $\rho_{c}$ $\times$ 10$^{-3}$ $\mu$$\Omega$ cm, the resistivity $\rho_{ab}$ may actually be smaller than the estimation.
Here the ratio $\rho_{ab}$(300K)/$\rho_{ab}$(4K) $\sim$ 40 is comparable to that of a bulk used in Ref.~\onlinecite{RMP}.
Hence we consider there is no degradation of the sample by the solvent. 
Figure \ref{resistivity} also shows a transition temperature of $T_c$ = 1.69 K and a broader transition temperature width of $\Delta$$T$ $\approx$ 200 mK. 
There was no decrease of the resistivity when magnetic field of 3000 G was applied parallel to the $c$ axis. 
Our sample shows no suppression of $T_c$ nor enhancement to 3K \cite{impurity,3K}.
Here the resistivity retained its flat tail below $T_c$.
The result shows the flow of vortices can be occurred by quantum fluctuations of the superconducting phase $\theta$ \cite{Haviland}.
The results show transport properties of the submicron Sr$_2$RuO$_4$ single crystals because the broader transition temperature width and the quantum fluctuations of the phase are characterized in mesoscopic superconductors \cite{degennes}. 

\begin{figure}[t]
\begin{center}
\includegraphics[width=0.49\linewidth]{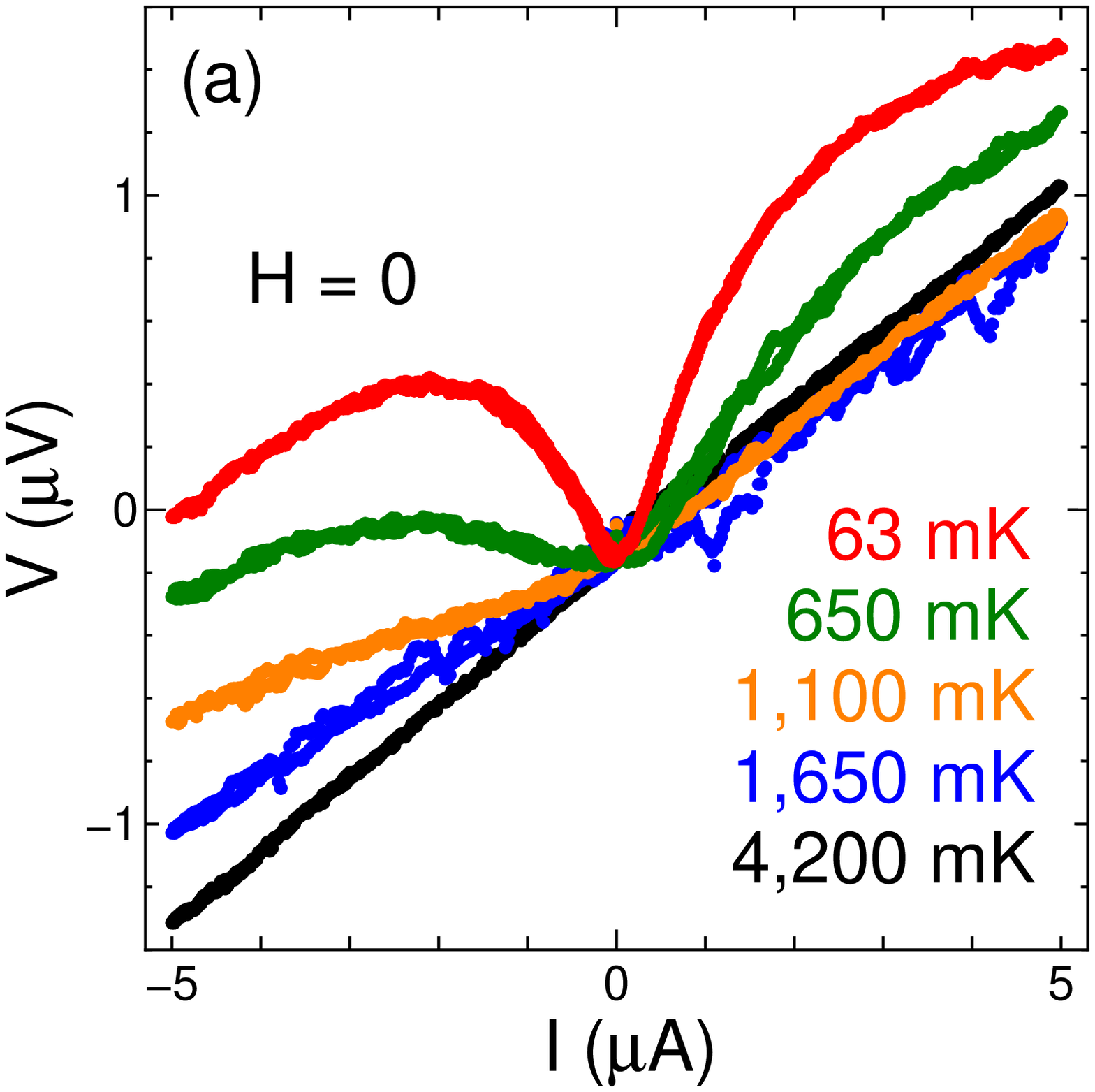}
\includegraphics[width=0.49\linewidth]{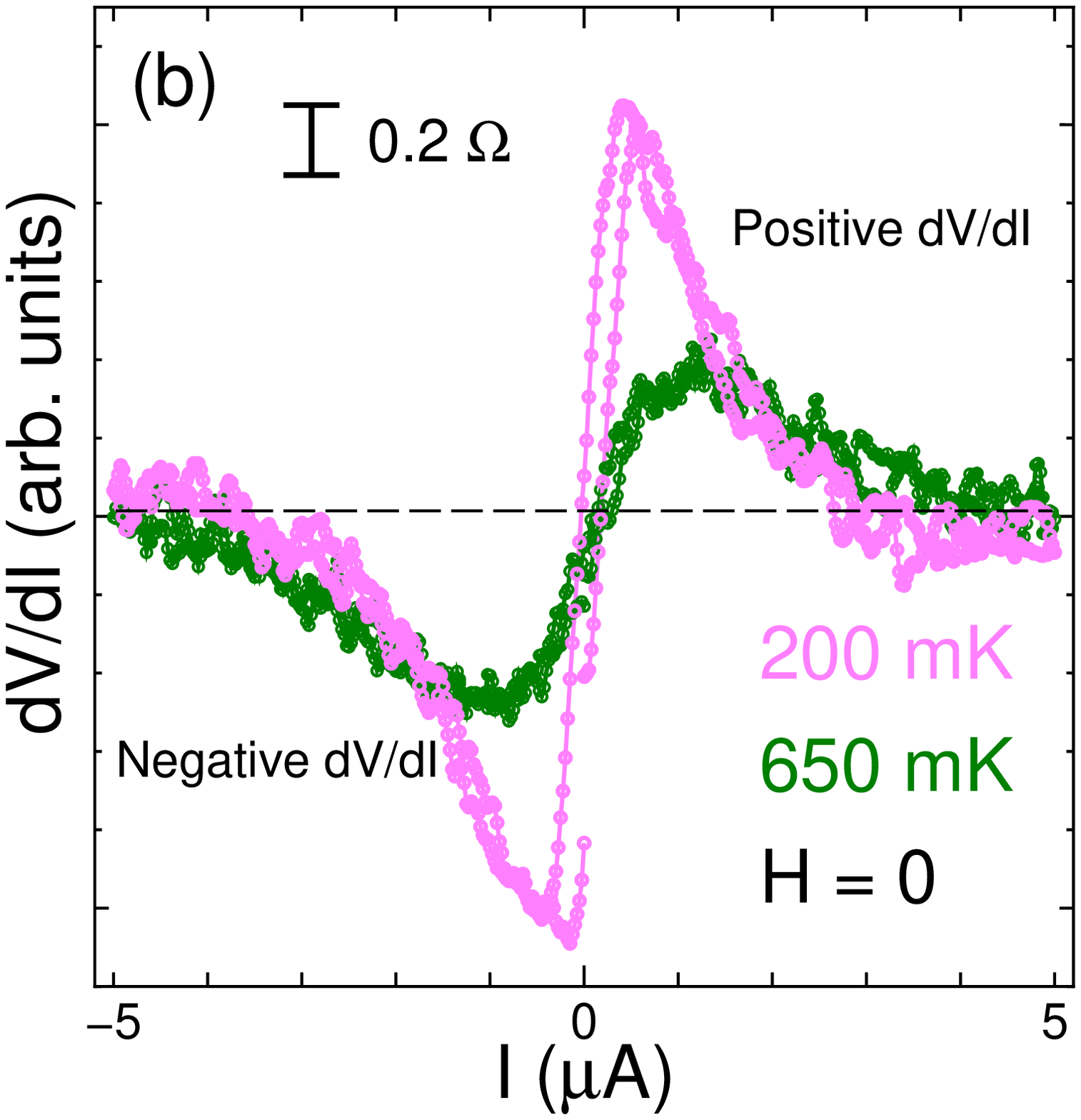}
\caption{
(a): Results of DC measurements. Voltage $V$ is plotted as a function of bias current $I$ in the absence of magnetic field for several choices of temperatures. The amplitude of the ANV increases with decreasing temperature and shows maximum below 200 mK. 
(b): Results of AC measurements. Differential resistance $dV/dI$ versus bias current $I$ is shown at $T$ = 200 mK and 650 mK. 
The upper and lower regions of the transverse dotted line represent positive and negative differential resistance, respectively.
As shown in (a) and (b), the parity violation are confirmed in two different measurements.
}
\label{IV}
\end{center}
\end{figure}

We observed anomalous the $I-V$ characteristics in zero magnetic field.
Figure \ref{IV}(a) shows $I-V$ curves at temperatures with typical flat tail resistances of $R^{*}\approx$ 0.16 $\Omega$.
In general, the voltage in $I-V$ curves for metals, quantum Hall systems and Josephson junction is always $odd$ function of bias current, which is a result of parity conservation.
Surprisingly, $V$ is not an $odd$ function of $I$ at all. 
In what follows, we define anomalous nonlinear voltage (ANV) as a component of measured voltage given by $even$ function of $I$.
The ANV implies the violation of parity.
The amplitude of the ANV increases with decreasing temperature in zero magnetic field and shows maximum below 200 mK.
In order to eliminate completely the possibility of instrument malfunction in the DC measurements, $I-V$ curves were measured with a micro-voltmeter (AM 1001, Ohkura Electric Co.) with a battery-powered current source. 
Furthermore, in the AC measurements, we also measured the differential resistance $dV/dI$ as a function of the bias current using lock-in techniques.  
Figure \ref{IV}(b) clearly shows $dV/dI$ has a $odd$ component of $I$.
The parity violation in the $I-V$ characteristics are confirmed in both the DC and AC measurements. 
Moreover, we confirmed that the anomalous effect was reproduced in several samples.

\begin{figure}[t]
\begin{center}
\includegraphics[width=0.6\linewidth]{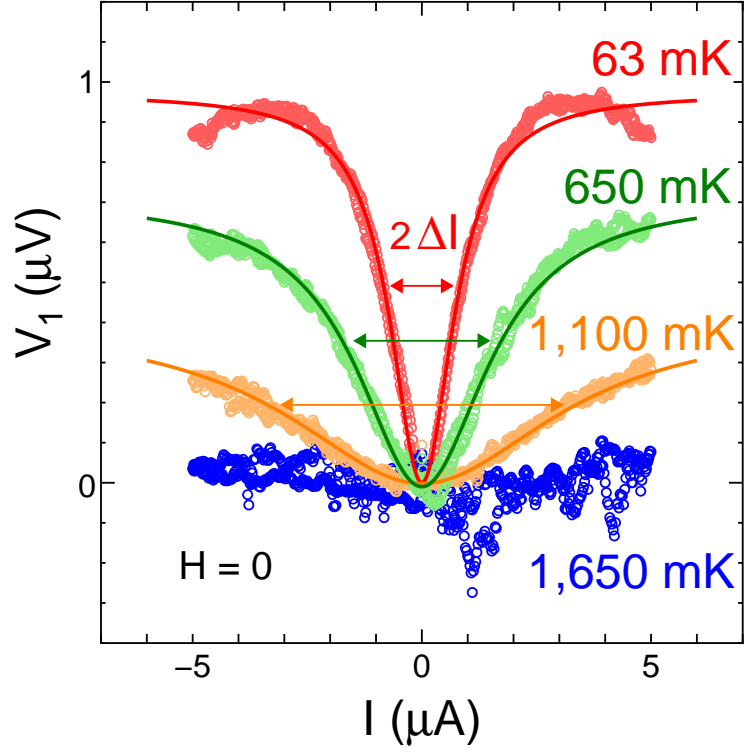}
\caption{
The anomalous nonlinear voltage (ANV) which is described by the voltage $V_1$ is extracted from $I-V$ characteristics in Fig. \ref{IV}, where the lines are Lorentzian fitting curves. 
The half-widths $\Delta$$I$ of the fitting curves are represented by two-headed arrows.
We eliminated the offset voltage of 0.13 $\mu$V in order to discuss the ANV.
}
\label{Lorentzian_temp}
\end{center}
\end{figure}

To analyze the ANV in more detail, we subtract linear part (ohmic contribution to voltage) from the $I-V$ curves in Fig. \ref{IV}(a).
The results are shown in Fig. \ref{Lorentzian_temp}. 
We clearly find that the ANV has symmetric with respect to the zero bias current, (i.e., $V_1(+I)=V_1(-I)$). 
Here the voltage $V_1$ represents the ANV of the induced voltage $V=R^*I+V_1$.
These curves are described well by Lorentzian curve as shown with lines. 

We discuss the physical difference between parity-violating $I-V$ characteristics and negative resistance. 
Negative resistance itself is an unusual phenomenon. The phenomenon, however, is possible. 
It is intuitive to compare our result to the negative resistance of mesoscopic charge density waves (CDW) reported in Ref.~\cite{zant}.
In their report, the negative resistance was attributed to the backflow of quasiparticle of CDW.  
This does not relate to parity violation of $I-V$ curves.
On the other hand, our observation of $V_1(+I)=V_1(-I)$ in the submicron Sr$_2$RuO$_4$ is the parity violation, which is a qualitatively different phenomenon from the CDW case.
The parity violation of $I-V$ curves must naturally include both negative resistance and negative differential resistance.
In addition, since the measurement was carried out with the four-terminal configuration, the observation does not violate any basic laws, such as energy conservation. 
From the reasons, we focus discovery of the parity violation.

What is the origin of the ANV in the $I-V$ characteristics?
The flat tail of resistivity (which often appears in superconductors owing to quantum fluctuations of the phase \cite{Haviland}) shown in Fig. \ref{resistivity} may suggest that the flow of vortices cause the ANV.
However, the dynamics of the usual Abrikosov vortex in type-I$\hspace{-.1em}$I superconductors cannot explain the parity violation in the $I-V$ curves.
Therefore we need to consider unconventional vortices characterized by $\vec{l}$ and $\vec{d}$ textures as in superfluid $^3$He-A. 
Here the $\vec{l}$ vector represents the direction of the pair angular momentum parallel to $c$ axis and the $\vec{d}$ vector describes the spin configuration of a pair. 
These internal degrees of freedom are characters of spin-triplet $p$-wave superconductivity.
Kerr effect measurements of Sr$_2$RuO$_4$ \cite{kerr} and Josephson tunneling measurements \cite{domain} respectively suggested chiral domain size to be 50 $\sim$ 100 $\mu$m and 1 $\mu$m. 
Recently the experiment for the 3-K phase of Sr$_2$RuO$_4$ also reveals that the domain size is $\sim$10 $\mu$m \cite{kambara}.
Since our sample electrode spacing is 0.63 $\mu$m, our sample is considered to have a single chiral domain.
The spin degree of freedom represented by $\vec{d}$ allows the formation of $\vec{d}$ textures in the single domain.
In bulk Sr$_2$RuO$_4$, the spin-orbit interaction favors the alignment of $\vec{d}$ and $\vec{l}$ in zero magnetic field. 
Knight shift measurement have recently suggested that a magnetic field $H$ $>$ 200 G may neutralize the interaction \cite{murakawa}. 
However, in submicron Sr$_2$RuO$_4$, quantum fluctuations of the phase disturb the alignment of $\vec{d}$ in a particular direction. 
Thus $\vec{d}$-textures would be possible in a single domain of Sr$_2$RuO$_4$.

In what follows, we consider a spin-triplet Cooper pair as a spin-1 Boson with charge $2e$.
Babaev theoretically predicted that the ground state of such Boson system can have magnetic 
spin textures $\vec{\boldsymbol{s}}$ characterized by a topological invariant known as helicity in zero magnetic 
field~\cite{babaev}. 
The equivalence between gauge transformation and spin rotation arises a term
\begin{eqnarray}
{B_{k}}^2=\left( -\frac{cM}{4e^2n}[\nabla_{i}J_{j}-\nabla_{j}J_{i}]+\frac{\hbar c}{4e}(\vec{\mbox{\boldmath $s$}} \cdot \nabla_{i}\vec{\mbox{\boldmath $s$}}\times \nabla_{j}\vec{\mbox{\boldmath $s$}})\right)^2
\label{eq:faddeev term}
\end{eqnarray}
in the Ginzburg-Landau energy functional,
where $\nabla_{i} = \frac{d}{dx_{i}}$, $\boldsymbol{J}$ is electric current,
$M$ is the mass of a Boson, and $n$ is the Boson density.
In Ref.~\onlinecite{toroidal}, the authors described a nontrivial topological 
structure in the simplest toroidal knot soliton.
The main distinction between knotted solitons in spin-triplet superconductors and the topological defects of $^3$He in Ref.~\onlinecite{mineev} is that the appearance of terms 
$\propto$ $(\vec{\mbox{\boldmath $s$}} \cdot \nabla_{i}\vec{\mbox{\boldmath $s$}}\times \nabla_{j}\vec{\mbox{\boldmath $s$}})^2$. 
This results in the knotted solitons being protected against shrinkage by an energy barrier.
According to Ref.~\cite{babaev}, the size of the knotted soliton is comparable to magnetic penetration length.
Our sample satisfies this requirement about the size.
We note that the sample size $\approx$ 1 $\mu$m and $\lambda_{ab}$ $\approx$ 152 nm.

\begin{figure}[t]
\begin{center}
\includegraphics[width=0.6\linewidth]{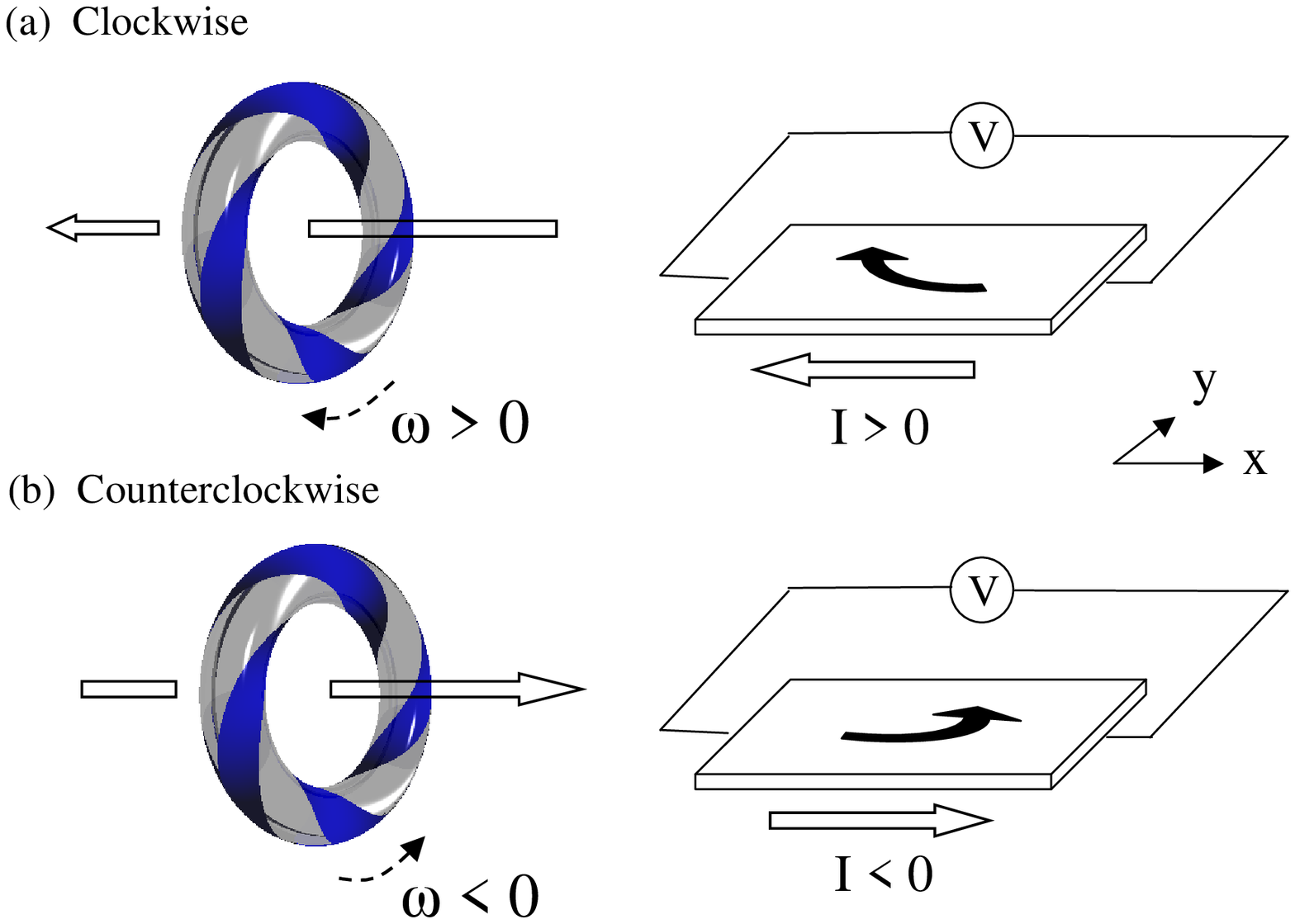}
\caption{
A model of vortex which violates parity.
A topological invariant features a helical structure of the vortex.
The blue spiral line on the torus represents a magnetic helical structure of
spin texture described by 
$(\vec{\mbox{\boldmath $s$}} \cdot \nabla_{i}\vec{\mbox{\boldmath $s$}}\times \nabla_{j}\vec{\mbox{\boldmath $s$}})$. 
Under the bias current, the spin texture around the torus moves periodically 
from the inside to the outside.
Open arrows represent the current flow $I$.
Solid arrows show the direction of the Magnus force. 
}
\label{model}
\end{center}
\end{figure}

In the Sr$_2$RuO$_4$ single domain, a helical vortex could be created by the spin degree of 
freedom. If we accept the existence of such helical vortices, 
the anomalous $I-V$ characteristics could be 
understood as a results of the conservation of a topological invariant in the helical vortex.
Now let us consider the \textit{hydrodynamics} of a helical vortex.
In the initial state of the bias current $I= 0$, the helical vortex does not move. 
When we switch on a bias current $I>0$, 
a clockwise helical vortex ($\omega > 0$) exerts the Magnus 
force in a direction perpendicular to the current as shown in Fig.~\ref{model}(a).
Here $+y$ indicates the direction of Magnus force.  
On the other hand, for a bias current $I < 0$, 
the helical vortex changes its rotation frequency from clockwise ($\omega > 0$) to counterclockwise 
($\omega < 0$) in order to retain topological invariant. 
As a consequence, the counterclockwise helical vortex also exerts
the Magnus force $+y$ direction as shown in Fig. \ref{model}(b). 
Thus the direction of Magnus force is independent of the direction of bias current.
According to the Josephson's equation, the motion of a vortex to the $+y$ direction
induces voltage across the sample of Sr$_2$RuO$_4$ in the $x$ direction. 
In this way, the presence of helical vortices explains the ANV.

Let us discuss the amplitude voltage of the ANV using a simple energy conversion formula $eV \sim \mu_{B}B$, where $\mu_{B}$ is the Bohr magneton, in the $I-V$ curves. 
The energy of amplitude voltage $V_1$ = 0.97 $\mu$V of the ANV at 63 mK is comparable to the energy of the magnetic field $H \approx 200$ G which neutralize the spin-orbit interaction in bulk Sr$_2$RuO$_4$ \cite{murakawa}. 
Thus we consider the amplitude voltage of the ANV may exhibit the contribution of the helical vortices.  

Finally we briefly discuss meaning of this experiment. 
Although Sr$_2$RuO$_4$ is a candidate of spin-triplet superconductor, 
this conclusion is still under debate. We show in this Letter that 
a Cooper pair in Sr$_2$RuO$_4$ has a spin degree of freedom. Thus
our results exhibit a possibility of spin-triplet symmetry \cite{ichioka}. 
When the helical vortices exist in a sample, such topological defects may
affect Hall conductivity. 
This infer stems from an analogy between the Chern-Simons term in quantum Hall effect and helical spin term in the Babaev's argument. 
Thus we believe that transport experiments which are sensitive to probe geometry would display more interesting phenomena reflecting internal degree of freedom of a Cooper pair.

In summary, we have observed an unconventional vortex which violates the parity in a single domain 
of Sr$_2$RuO$_4$ using a transport measurement. 
The $I-V$ characteristics of submicron Sr$_2$RuO$_4$ shows that the voltage has anomalous components which are \textit{even} function of the bias current.
We consider a vortex with a helical internal structure characterized by a Hopf invariant. The invariant of vortex is protected while the vortex is moving under the bias current.
By a simple argument, we show that the hydrodynamics of the helical vortex causes the anomalous $I-V$ characteristics to retain the topological invariant.

\begin{acknowledgements}
The authors thank A. Ishii, S. Tsuchiya, T. Toshima, T. Matsuura, T. Kuroishi, and D. Meacock for experimental help and useful discussions. 
We also are grateful to Y. Maeno, and T. Matsuyama. 
H.N. acknowledges support from the Japan Society for the Promotion of Science. 
\end{acknowledgements}

\newpage

\begin{thebibliography}{0}
\bibitem{parity} Here we apply a simple argument as follows. 
Both electric field $E$ (voltage drop $V$ per unit length) 
and electric current $I$ have odd parity, 
(i.e., change their signs under the space inversion). 
Thus a relation $V= \sum_{n=1}^{\infty} a_n I^{2n+1}$ is usually expected.
However when a relation $V= \sum_n a_n I^{2n+1} + \sum_m b_m2 I^{2m}$ holds with $b_m\neq 0$, 
$V$ and $I$ cannot have odd parity at the same time.
In this paper, we claim that some defect (vortex) of parity nonconservation 
causes such anomalous $I-V$ characteristics.
\end{thebibliography}
\end{document}